# DIPOLE EMISSION IN FINITE PHOTONIC BAND-GAP STRUCTURES: AN EXACTLY SOLVABLE ONE-DIMENSIONAL MODEL

Jonathan P. Dowling *

*Quantum Technologies & Algorithms Group*
*Information & Computing Technologies Research, Section 365*
*NASA Jet Propulsion Laboratory, California Institute of Technology*
*MS 126-347, 4800 Oak Grove Drive, Pasadena, California 91109-8099*

## ABSTRACT

I consider an exact model of atomic spontaneous dipole emission and classical dipole radiation in a finite photonic band-gap structure. The full 3D or 2D problem is reduced to a finite 1D model, and then this is solved for analytically using algebraic matrix transfer techniques. The results give insight to the electromagnetic emission process in periodic dielectrics, quantitative predictions for emission in 1D dielectric stacks, and qualitative formulas for the 2D and 3D problem.

## 1.    INTRODUCTION

From the beginning, one of the principal potential applications of photonic band-gap (PBG) materials was to the control of atomic spontaneous dipole emission [1, 2]. In particular, it was predicted that emission could be suppressed for dipoles located inside the PBG material when their resonant emission frequency $\omega_0$ was in the photonic band gap. In this frequency range the electromagnetic density of modes $\rho(\omega)$ is very small. Resonant enhancement of emission was expected at the photonic band edges where the density of modes (DOM) was anomalously large. These phenomena are a result of the so-called *Purcell Effect*. Purcell concluded that nontrivial boundary conditions on the electromagnetic field surrounding a dipole emitter can alter the emission rate. They do so by altering the DOM and the electromagnetic modal field $e(\omega_0, r_0)$ at the position $r_0$ of the dipole [3]. (This original paper of Purcell was actually a short abstract for a talk to be given at an annual meeting of the American Physical Society. He apparently never wrote up anything more about it nor even gave the talk. Nevertheless, he gives in this short abstract the first quantitative formula for the alteration of spontaneous emission in cavities.)

In 1992, Bowden and I gave a general formalism for computing atomic or classical dipole radiation rates in an inhomogeneous dielectric medium, of which a PBG material is a specific example [4]. Although our theory was completely general, it takes a supercomputer to figure out the DOM and electromagnetic modal structure of a full 2D or 3D PBG structure. Hence, after presenting the general formalism, we gave an exactly solvable model in terms of an infinite 1D Kronig-Penney model of the full 2D or 3D structure. This approximation is due to John and Wang. It amounts to replacing the Brilloun zone (BZ) of the 2D structure by a perfect circle, or of the 3D one by a perfect sphere—the same circle or sphere for both polarizations [5]. Physically, this model is in the spirit which Yablonovitch searched for a full 3D PBG by experimentally investigat-ing structures whose BZ was as close to spherical as possible [5]. Hence, this model gives qualitative predictions of what to expect in 2D and 3D, and quantitative predictions in 1D periodic dielectric PBG slabs. Unfortunately, due to the low-dimensionality of the problem, as well as the assumption of an infinite 1D lattice, the DOM of this model is formally infinite at the edges of the photonic band gap. Since the spontaneous emission rate is proportional to this DOM, the model breaks down at the photonic band edge, producing unphysical divergent results.

In 1995 on a supercomputer, Suzuki and Yu used the general 3D formalism developed by Bowden and me. They were able to compute the emissive power of a point dipole embedded in an infinite 3D fcc dielectric PBG lattice [7]. Using the same approach, working with Schultz and collaborators, they also were able to accurately model the emission of a radio-wave dipole oscillator embedded in a 2D array of dielectric rods. They obtained quantitative agreement between theory and experiment [8]. This work conclusively demonstrated the utility of our approach in realistic experiments. However, the need for numerically intensive computations obstructs intuitive understanding of the basic physics involved. For this reason, I have gone back to the old one-dimensional Kronig-Penny model of my 1992 paper with Bowden and removed the restriction of the *infinite* 1D lattice assumption. Perhaps surprisingly, the restriction to finite periodic arrays makes the analytic calculation much harder, due to the fact that the system is not absolutely periodic. This lack of symmetry greatly complicates the algebra, but a systematic attack using techniques of matrix-transfer theory has yielded a solution that I will present here.

The first phase of the calculation has already been carried out in a previous work by Bendickson, Scalora, and me. In this paper we were able to find an exact expression for the DOM of a large class of an arbitrary, 1D PBG structures, using a matrix transfer approach [9, 10]. This technique was also applied by us numerically in order to model the spontaneous emission rate of active emitters in an experiment conducted by our group. We considered emitters near the photonic band edge of a 1D semiconductor superlattice [11]. We were also able to use these ideas in order to understand theoretically and experimentally the anomalously large group delay for ultra-short optical pulses propagating at photonic band-edge frequencies in 1D PBG structures [12, 13]. The solution $\rho(\omega)$ of the DOM follows from the eigenvalue solutions of the 1D Helmholtz equation in an inhomogeneous medium. Discovering the electromagnetic modes $e(\omega_0, r_0)$ amounts to the logical next step of finding the eigenvectors, which is usually a more difficult problem. It is this final step that I wish to present here and then enfold into the previous results to give a general solution to the finite, 1D, spontaneous emission problem in a periodic dielectric.

In Sec. 2, I will review the general theory of dipole emission in inhomogeneous dielectrics and the consequences of a restriction to 1D. In Sec. 3, I will develop the analytic matrix transfer techniques for a finite 1D PBG structure and from this result extract analytic expressions for the DOM and electric modal fields. From these formulas, according the Purcell prescription, I can finally give a result for atomic emission in a finite 1D model that is free from band-edge singularities. In Sec. 4, I will apply this general theory to the special case of emitters in a traditional quarter-wave stack as a simple example. Then in Sec. 5, I will summarize and conclude.

In addition to the applications to atomic emission, this work contains for the first time the general, exact solution for the modal fields $e(\omega_0, r_0)$ in a 1D, finite periodic structure. This solution should have many applications to the study electromagnetic wave propagation in 1D periodic media.

## 2. SPONTANEOUS EMISSION IN INHOMOGENEOUS MEDIA

### 2.1 Fermi's Golden Rule and the Purcell Effect

It has been known for some time now that the effect of a cavity on emission rates of atoms is essentially classical [14–16]. This can be seen by considering Fermi's golden rule. Suppose we have a single two-level atom coupled to the electromagnetic field. Further suppose that the atom is in an initial excited state $|i\rangle$ and the field contains no photons. The state of the system can be written $|i, 0_{\boldsymbol{k}}\rangle$ where $0_{\boldsymbol{k}}$ indicates the lack of photons of wave number $\boldsymbol{k}$. Let the final state of the system consist of the atom in some final state $|f\rangle$ after the release of a photon. The final state of the system is then $|f, 1_{\boldsymbol{k}}\rangle$. Fermi's golden rule yields for the spontaneous transition rate $w_{fi}$ gives the following,

$$ w_{fi} \;=\; \frac{2\pi}{\hbar}\;\rho\big(\hbar\omega_0\big)\;\big|\langle f, 1_{\mathbf{k}}|\boldsymbol{\mu}\bullet\mathbf{e}(\omega_0,\mathbf{r}_0),|i, 0_{\mathbf{k}}\rangle\big|^2 \;, \qquad (1) $$

where $\hbar$ is Dirac's constant (Planck's constant divided by $2\pi$) and $\boldsymbol{\mu}$ is the dipole moment operator for the atomic transition. It turns out that the DOM is the same, both from a QED and a classical standpoint [4]. The expectation of the quantum mechanical dipole moment $|\langle f|q\,\boldsymbol{x}|i\rangle|^2$ can be identified with the time averaged dipole moment of a classical oscillator with $q$ the charge and $\boldsymbol{x}$ the charge displacement. At the single photon event level the electric field operator $\boldsymbol{e}(\omega_0, \boldsymbol{r_0})$ has two equal contributions from the atomic radiation-reaction (self) field and from the QED electromagnetic vacuum fluctuations. Both contributions experience the same boundary conditions, so they exhibit the same cavity QED correction factor [17]. For a single atomic spontaneous emission event, they both contribute equally to the transition rate, Eq. (1). However, in the classical correspondence-principle limit of large quantum numbers and many photons, only the radiation-reaction term contributes. The effect of the single vacuum photon is negligible in this classical regime. Hence, the classical calculation will be smaller than a full quantum calculation by only a factor of two, which is apparent at the single quantum level. (Purcell's original formula is too small by a factor of two— indicating that he had a classical derivation in mind in 1945, before QED was invented.) For a single atom, one can think of the spontaneous emission as being stimulated by equal parts radiation-reaction and vacuum fields. With these considerations the classical expression becomes,

$$ P(\omega_0, \mathbf{r}_0) \;=\; 2\pi\kappa\rho(\omega_0)\;\big|\boldsymbol{\mu}\bullet\mathbf{e}(\omega_0, \mathbf{r}_0)\big|^2 \;, \qquad (2) $$

where $\hbar w_{fi} \to P(\omega_0, \mathbf{r}_0)$ is the classical Poynting vector power output in the limit of large quantum numbers and large numbers of photons. Here, $\boldsymbol{\mu}$ is the classical dipole moment and $\mathbf{e}(\omega_0, \mathbf{r}_0)$ the electric modal function evaluated at the dipole frequency and position. The factor $\kappa$ is a dielectric enhancement factor that depends on the value of the index of refraction $n(\omega_0, \mathbf{r}_0)$, also at the frequency and position of the dipole. For a tenuous medium such as a gas the dielectric constant has the effect of concentrating the electric field at the dipole position by a factor of $\kappa = n^3$. There is one factor of $n$ for each modal degree of freedom. However, if the dipole is embedded in a very dense material such as a dielectric solid then local field effects come into play and there is a different factor, namely $\kappa = (2+n^2)/3$ in that case [18].

The electric field modal function is normalized such that [4, 19],

$$\int\limits_V n^2(\mathbf{r}) \left| \mathbf{e}(\mathbf{r}) \right|^2 dV = 1 , \tag{3}$$

where $V$ is the cavity mode volume. Physically, this gives the modal field in units that normalizes the electromagnetic energy to unity. The time-dependent decay of this energy is proportional to the $Q$ of the cavity, and allows us to arrive at Purcell's formula in terms of cavity $Q$ and the volume $V$ of the mode [3].

When comparing the effect of the cavity on the enhancement and suppression of dipole emission, it is necessary to choose a proper control system. For example, it would not be fair to embed a dipole in a high-index region of a dielectric lattice, and then compare the emission with that of free space. The dielectric local-field enhancement factor alone would provide a somewhat trivial enhancement apart from the cavity-induced interference effects which we are interested in. For this reason, one should strive to use as a control the emission rate of a dipole embedded in an infinite homogenous medium of index $n(\mathbf{r}_0)$. For that is the value of the inhomogeneous index at the dipole location in the photonic band structure. This is practicable from an experimental point of view when one can scale the emission spectrum data from the PBG sample by that of a reference emitting bulk medium to obtain a normalized power emission [11],

$$p(\omega_0, \mathbf{r}_0) = \rho(\omega_0) \left| \hat{\boldsymbol{\mu}} \bullet \mathbf{e}(\omega_0, \mathbf{r}_0) \right|^2 , \tag{4}$$

where $\hat{\boldsymbol{\mu}} = \boldsymbol{\mu}/|\boldsymbol{\mu}|$ is the normalized dipole moment unit vector. Hence, Eq. (4) represents the pure geometrical effect of cavity-induced interference on the dipole emission. All physical factors, such as the local-field factor and absolute dipole moment, have been scaled out. Since $\hat{\boldsymbol{\mu}}$ is a fixed vector, we need only now compute $\rho(\omega_0)$ and $\mathbf{e}(\omega_0, \mathbf{r}_0)$ to give the final emission power.

## 2.2    One-Dimensional Scattering and the Wave Equation

The formulas given above for calculating the bulk-scaled emission rate $p(\omega_0, \mathbf{r}_0)$, require the functional form of the density of modes $\rho(\omega)$ and the modal functions $\mathbf{e}(\omega_0, \mathbf{r}_0)$. For a 1D problem, this reduces to solving the 1D Helmholtz wave equation [4],

$$y''(x) + \frac{\omega^2}{c^2} n^2(x) y(x) = 0 , \tag{5}$$

where I have neglected the dispersion of the index function $n(x)$, as is usually done. The solution of this eigenvalue problem yields the dispersion relation $\omega = \omega(k)$ and the modal eigenvectors $y = y_k$, as a function of the eigenvalue (wave number) $k$. Differentiating the dispersion relation gives the density of modes [4],

$$\rho(w) \; = \; \frac{dk}{dw} \; , \tag{6}$$

which is the reciprocal of the group velocity [9, 12, 13]. For the dipole emission problem, Eq. (6) can be interpreted as counting the number of modes $\Delta k$ available for photons to radiate into, per unit frequency $\Delta\omega$. The more of these there are, the faster the dipole radiates. The eigenvectors $y_k$ can then be normalized as per Eq. (3), giving all the pieces for the solution, Eq. (4). However, even in 1D this is a nontrivial problem to carry out analytically. Eq. (5) is only exactly solvable for a small class of functions $n(x)$. Of course the equation may be solved numerically by finite difference techniques, but here I take a different tack and use methods from 1D scattering theory [9, 13].

Consider Fig. (1), where I treat the inhomogeneous 1D dielectric in terms of a scattering problem. Assuming an incoming field of unit amplitude from the left and zero field from the right, the scattered fields are the complex transmitted and reflected field amplitudes $t$ and $r$. Under the assumptions that the index $n(x)$ is linear, lossless, and dispersionless, this scattering process may be written in transmission matrix form as [9,13],

$$\boldsymbol{\lambda} = \begin{bmatrix} 1 \\ r \end{bmatrix} \; = \; \begin{bmatrix} 1/t & r^*/t^* \\ r/t & 1/t^* \end{bmatrix} \begin{bmatrix} t \\ 0 \end{bmatrix} \equiv \hat{\mathbf{M}} \boldsymbol{\rho} \; , \tag{7}$$

where $\boldsymbol{\lambda}$ and $\boldsymbol{\rho}$ are the left and right boundary-condition vectors, respectively, and $\hat{\mathbf{M}}$ is the transfer matrix. The form of $\hat{\mathbf{M}}$ in Eq. (7) is completely determined from linearity and time-reversal symmetry [9, 13]. If the further condition of losslessness is imposed, then in addition we have $\det |\hat{\mathbf{M}}| = 1$. The scattering amplitudes can be written in terms of amplitude and phase as, $t = e^{i\varphi}\sqrt{T}$ and $r = e^{i\psi}\sqrt{R}$, where, if energy is conserved, then the relation $T + R = 1$ holds. In general the phases $\varphi$ and $\psi$ are not equal. However, if the index profile is symmetric, $n[-(x-d/2)] = -n[x-d/2]$, then we have $\varphi - \psi = \pm\pi/2 \mod 2\pi$, which is a well-known property of symmetric beam splitters. Here, this is just a consequence of parity conservation (reciprocity) in a generally symmetric structure [13]. Writing $\varphi = kd$, where $d$ is the physical length of the structure, I can now relate $k=k(\omega)$ to the solution of the scattering problem. In other words, given $n(x)$, if we can solve for $t$ and $r$ as functions of frequency $\omega$, we can then extract $k = \varphi(\omega)/d$. Finally, differentiating this expression gives us the density of modes $dk/d\omega$, Eq. (6). If I write the transmitted amplitude in Argand notation, $t = u + iv$, then the DOM can be written as [9, 13],

$$\rho(\omega) \; = \; \frac{1}{d} \; \frac{v'u - vu'}{v^2 + u^2} \; , \tag{8}$$

where differentiation is with respect to frequency $\omega$. Thus the DOM is extracted from the solution to the scattering problem. To get out the unnormalized modal eigenvector, $\boldsymbol{E}(\omega_0, \boldsymbol{r_0})$, let us recall that the general solution to the Helmholtz Eq. (5) can be written as $y(x) = A f(x) + B g(x)$. Here, $f$ and $g$ are the two independent solutions required for a second order differential equation. If $n(x)$ has an exactly solvable form, then $f$ and $g$ are known. For example, if *n(x)* is constant, then *f* and *g* are sines and cosines, or complex exponentials. If $n^3(x) = ax+b$ is a linear function, then *f* and *g* have solutions in terms of the Airy functions Ai and Bi. Taking *f* and *g* to be known, I may apply the boundary conditions (BC) of Eq. (7) and solve for complex *A* and *B* in terms of complex *t* and *r*. This is done by setting the sum of the BCs at the left interface equal to the sum of the components of *E(x)* evaluated at the left boundary. A similar equation holds for the right boundary. Solving the two equations simultaneously for the two unknowns *A* and *B*, yields

$$A = \frac{tg(0) - (1 + r)g(d)}{f(d)g(0) - f(0)g(d)} \quad , \tag{9a}$$

$$B = \frac{(1 + r)f(d) - tf(0)}{f(d)g(0) - f(0)g(d)} \quad , \tag{9b}$$

which allows us to write as *E(ω,x) = A(ω) f(ω,x) + B(ω) g(ω,x)*, recalling that *t=t(ω)* and *r=r(ω)*. If *f* and *g* are can not be found exactly, then they can be found numerically by either finite difference methods or by subdividing the index profile *n(x)* up into partitions. Over each partition one takes $n(x_i)$ to be constant and then applies a numerical matrix-transfer approach to solve for $E(\omega, x_i)$ in each subdivision [11].

Two further conditions are needed to make sure the solution is unique, and they are energy conservation, *T + R = 1*, and energy normalization, which can be done by first computing the modal energy as per Eq.(3), $U = \int_0^d n^2(x) \left| E(x) \right|^2 dx$, and then defining the eigenvector $e(w,x) = E(w,x) / \sqrt{U}$. This prescription gives us the 1D scaled emission rate, Eq. (4), as $p(\omega,x) = \rho(\omega) \left| e(\omega,x) \right|^2$, where I take the dipole moment perpendicular to the *x* direction. Physically, this corresponds to the rate at which the photons are emitted into the two modes in the ±*x* direction of a 1D structure.

## 2.    EMISSION IN FINITE 1D PHOTONIC BAND-GAP STRUCTURES

In this section, I will define a 1D PBG structure and derive properties of the N-period PBG stack in terms of properties of the unit-cell problem. The unit-cell wave equation (5) is assumed to be solved as was described in the previous section. Once this is done, I can then express the emissive properties of a dipole embedded in the stack in terms of quantities associated with the unit cell.

### 2.1    Properties of the 1D Stack

From Fig. 2, we see that a finite, 1D, *N*-period, quasi-periodic PBG structure can be defined by just repeating the unit cell *N* times. The new complex transmission and reflection coefficients are $t_N = e^{i\varphi_N} \sqrt{T_N} = u_N + iv_N$, and $r_N = e^{i\psi_N} \sqrt{R_N}$. From linearity, we have that the transfer matrix for the PBG stack $\hat{\mathbf{M}}_N$ can be written in terms of $\hat{\mathbf{M}}$ for the unit cell as $\hat{\mathbf{M}}_N = \hat{\mathbf{M}}^N$. I now wish to derive a simplified form of $\hat{\mathbf{M}}_N$ that allows us to find an exact functional form for the needed DOM $\rho_N(\omega)$ which is a global property of the *N*-period stack. We also need $e_n(\omega,x)$, a local function in the $n^{th}$ unit cell where the emitter is located. To find this, first note that the unit cell matrix $\hat{\mathbf{M}}$ has an eigenvalue equation

$$\varepsilon^2 - 2\varepsilon\Re\{1/t\} + 1 = 0, \tag{10}$$

where $\varepsilon$ is the eigenvalue, $\Re$ the real-part function, and $t$ is the complex transmission amplitude for the unit cell. This equation has two solutions $\varepsilon_\pm$ that are related by $\varepsilon_+\varepsilon_- = \det|\hat{\mathbf{M}}| = 1$, from energy conservation. Now I can relate $\Re\{1/t\}$ to the Bloch phase $\beta$ of the *infinite* periodic structure, corresponding to (hypothetically) continuing the lattice in Fig. 2 to infinity in both directions. From the definition of eigenvector, the Bloch vectors $\boldsymbol{b}$ obey the particular equation

$$\hat{\mathbf{M}}\mathbf{b} = \varepsilon_\pm^B \mathbf{b} = e^{\pm i\beta}\mathbf{b}, \tag{11}$$

that is, the vectors vary only in phase and not amplitude from cell to cell in the *infinite* structure, where $\varepsilon_\pm^B = e^{\pm i\beta}$ are the Bloch eigenvalues. Since Eq. (10) holds for all eigenvalues, it holds for $\varepsilon_\pm^B$, which yields the very important relation that $cos\beta = \Re\{1/t\}$, connecting the transmission amplitude to the Bloch phase, as promised. Important to note is that the Bloch phase depends only on properties of the unit cell. I am now ready to derive a simplified form of $\hat{\mathbf{M}}_N$.

Recall from the Cayley-Hamilton theorem that every matrix obeys its own eigenvalue equation [20], hence

$$\hat{\mathbf{M}}^2 - 2\hat{\mathbf{M}}\cos\beta + \hat{\mathbf{1}} = 0, \tag{12}$$

where $\hat{\mathbf{1}}$ is the two-by-two identity matrix. Then by mathematical induction [9], it is easy to establish the *Scattering Matrix Reduction Formula* (SMRF), namely,

$$\hat{\mathbf{M}}_N = \hat{\mathbf{M}}^N = \frac{\sin N\beta}{\sin\beta}\hat{\mathbf{M}} - \frac{\sin(N-1)\beta}{\sin\beta}\hat{\mathbf{1}}, \tag{13}$$

which allows us to express the transfer matrix of the entire $N$-period PBG stack in simple closed form in terms of the matrix of the unit cell $\hat{\mathbf{M}}$, Eq. (7), and simple trigonometric functions of the Bloch phase $\beta$. Every quantity on the right-hand-side of Eq. (13) depends only on properties of the unit cell, except for the explicit integer $N$ in the argument of the sine functions. I first define an auxiliary function [21], which is related to the Chebyshev polynomial of the first kind by $\Xi_N(\beta) = sin\,N\beta\,/\,sin\,\beta$. Then from Eq. (13) and the definition of $\hat{\mathbf{M}}$ in Eq. (7), I can solve for $t_N$ and $r_N$ implicitly to get,

$$\frac{1}{t_N} = \frac{1}{t}\,\Xi_N(\beta) - \Xi_{N-1}(\beta), \tag{14a}$$

$$\frac{r_N}{t_N} = \frac{r}{t}\,\Xi_N(\beta), \tag{14B}$$

which allows us to compute $T_N$, $R_N$, $\varphi_N$, and $\psi_N$, the details of which can be found in Refs. 9 and 13. In particular, in these papers we discussed the evolution of the gap, transmission and reflection in the gap and at the band-edge resonances, and the DOM and group delay in the gap and at the band edge—all in terms of analytic formulas.

## 2.2    Emission in the 1D Stack

Eqs. (14) for $t_N$ and $r_N$ can be used directly in Eq. (8) for the density of modes, allowing us to write it in closed form as,

$$\rho_N(\omega) = \frac{\frac{1}{2}\Xi_{2N}(\beta)\left[\nu' + \mu\mu'\nu\csc^2\beta\right] - N\mu'\nu\csc^2\beta}{Nd\left[\Theta_N^2(\beta) + \nu^2\Xi_N^2(\beta)\right]} \;,\tag{15}$$

where $Nd$ is the physical length of the stack, and $\Theta_N(\beta) = cos\,N\beta$ is an additional Chebyshev function that I have defined, which is related to that of the second kind [21]. I have defined the normalized real and imaginary parts of the unit cell transmission amplitude t as $\mu = \Re(t)/T = cos\,\beta$ and $\mu = \Im(t)/T$. We now have the first piece of the 1D projection of Eq. (4) for the normalized 1D dipole emission power, the DOM. We now need the modal field in the $n^{th}$ unit cell, normalized to the electromagnetic energy in the entire stack, $e_n\,(\omega,x) = E_n(\omega,x)\,/\,U_N$.

Looking at Fig. 2, we can see that the boundary conditions on the outer edges of the PBG stack is given in terms of $t_N$ and $r_N$. These are in turn simple functions of the unit cell scattering coefficients $t$ and $r$, as per Eqs. (14). Hence, what is needed is a way to propagate these outermost boundary conditions (BC) inwards, to give the correct BC on the $n^{th}$ unit cell in the stack, containing the dipole emitter. The electric field in the nth cell will have the general form, $E_n(\omega,x) = A_n(\omega)\,f(\omega,x) + B_n(\omega)\,g(\omega,x)$. Here, $f$ and $g$ are the independent solutions of the Helmholtz Eq. (5), assumed to be known and independent of $n$ or $N$. The constants $A_n$ and $B_n$ do not depend on the position $x$, and can be solved for as per Eqs. (9), once the $n^{th}$ cell boundary conditions are specified. To accomplish this, I will employ the *Scattering Matrix Reduction Formula (SMRF)*, Eq. (13). First, realize that the matrix-transfer equation for the $N$-period stack can be written as $\boldsymbol{\lambda}_N = \hat{\mathbf{M}}^N\boldsymbol{\rho}_N$, where $\boldsymbol{\lambda}_N = \begin{bmatrix} 1 \\ r_N \end{bmatrix}$ and $\boldsymbol{\rho}_N = \begin{bmatrix} t_N \\ 0 \end{bmatrix}$ define the BC at the outer edges. This equation can be rewritten as $\boldsymbol{\lambda}_N = \hat{\mathbf{M}}^{n-1}\hat{\mathbf{M}}\hat{\mathbf{M}}^{N-n}\boldsymbol{\rho}_N$, which takes us in to the nth unit cell. Hence the BC on the right-hand-side (RHS) of the nth cell is given by $\boldsymbol{\rho}_n = \hat{\mathbf{M}}^{N-n}\boldsymbol{\rho}_N$, and that on the left side by $\boldsymbol{\lambda}_n = \hat{\mathbf{M}}^{N-n+1}\boldsymbol{\rho}_N$. Hence, I have matrix equations for the right and left BC at the nth unit cell, which I can simplify by applying the SMRF to get, respectively (written in transpose form to conserve space),

$$\boldsymbol{\rho}_n^\dagger = \left[\frac{t^2\Xi_{N-n-1}(\beta) - \Xi_{N-n}(\beta)}{t^2\Xi_{N-1}(\beta) - \Xi_N(\beta)} \;,\; \frac{-r^2\Xi_{N-n}(\beta)}{t^2\Xi_{N-1}(\beta) - \Xi_N(\beta)}\right],\tag{16a}$$

$$\boldsymbol{\lambda}_n^\dagger = \left[\frac{t^2\Xi_{N-n}(\beta) - \Xi_{N-n+1}(\beta)}{t^2\Xi_{N-1}(\beta) - \Xi_N(\beta)} \;,\; \frac{-r^2\Xi_{N-n+1}(\beta)}{t^2\Xi_{N-1}(\beta) - \Xi_N(\beta)}\right],\tag{16b}$$

all in terms of known unit cell quantities $t$, $r$, and $\beta$. This provides an analytic expression for the BC at the edges of the $n^{th}$ unit cell. The field inside this cell has the form $E_n(\omega,x) = A_n(\omega)\, f(\omega,x) + B_n(\omega)\, g(\omega,x)$, where $f$ and $g$ are known properties of the unit cell wavefunction solution, but $A_n$ and $B_n$ are to be determined. To find these I follow the same procedure leading to Eq. (9) for the unit cell, but now with the BC of Eq. (16) to get,

$$A_n = \frac{-g[dn]\big(t^2\Xi_{N-n-1} - (1+r^2)\Xi_{N-n}\big) + g[d(n-1)]\big(t^2\Xi_{N-n} - (1+r^2)\Xi_{N-n+1}\big)}{(f[dn]g[d(n-1)] - f[d(n-1)]g[dn])(t^2\Xi_{N-1} - \Xi_N)}, \tag{17a}$$

$$B_n = \frac{f[dn]\big(t^2\Xi_{N-n-1} - (1+r^2)\Xi_{N-n}\big) - f[d(n-1)]\big(t^2\Xi_{N-n} - (1+r^2)\Xi_{N-n+1}\big)}{(f[dn]g[d(n-1)] - f[d(n-1)]g[dn])(t^2\Xi_{N-1} - \Xi_N)}, \tag{17b}$$

which, when combined with $E_n(\omega,x) = A_n(\omega)\, f(\omega,x) + B_n(\omega)\, g(\omega,x)$, provides for the first time (to my knowledge) an exact solution to the 1D wave equation in a general, finite, periodic structure. Without the explicit form of the independent solutions to the unit cell wave equation, $f$ and $g$, electric field $E_n$ can not be further simplified.

Note that the $n^{th}$ cell electric field, as defined implicitly in Eqs. (17), is not normalized properly for the spontaneous emission problem. It is currently normalized such that the value of the right-moving field at the left edge of the PBG stack has amplitude one and zero phase. This is a standard normalization for the scattering problem but not for the spontaneous emission formula. Hence, given $E_n$ above, I define the total energy of the $N$-period stack as,

$$U_N(\omega) = \sum_{n=1}^{N} U_n(\omega) = \sum_{n=1}^{N} \int_{(n-1)d}^{nd} n^2(x)\big|E_n(\omega,x)\big|^2 dx, \tag{18}$$

which allows us then to find the correct normalized eigenmodes functions in the $n^{th}$ cell as,

$$e_n(\omega,x) = \frac{E_n(\omega,x)}{\sqrt{U_N(\omega)}} \; . \tag{19}$$

This is now in the correct form for use in the spontaneous emission formula, Eq. (4), which when taken with the density of mode formula, Eq. (15), yields the general solution for the dipole emission problem in this 1D model. In the next section I will illustrate how this all works with the specific example of a quarter-wave stack.

## 3.   EMISSION IN A QUARTER-WAVE STACK

In this section I will take the general theory developed in the previous section and apply it to the simple specific example dipole emission in a finite, 1D, quarter-wave stack. The quarter-wave unit cell and $N$-period stack are depicted in Figs. (3a) and (3b), respectively. For simplicity, I choose as the unit cell a two-layer region of constant, dispersionless indices $n_1$ and $n_2$ of lengths $a$ and $b$, respectively. There is a real $n_1$-to-$n_2$ interface at $x = a$, and a real $n_2$-to-$n_1$ interface at $x = d = a+b$, connecting the $n_2$ layer to the semi-infinite $n_1$ region on the right. At $x = 0$ there is a virtual interface connecting the $n_1$ region of the unit cell to the $n_1$ semi-infinite region on the left. When this unit cell is repeated $N$ times, it generates the quarter-wave stack in Fig. (3b), also surrounded by an infinite $n_1$ region. The quarter-wave condition requires that the $n_1$ and $n_2$ layers have an optical thickness that is a quarter of some reference wavelength $\lambda_0$, which requires that $n_1a = n_2b = \lambda_0/4 = \pi c/(2\,\omega_0)$, where $\omega_0 = 2\pi c/\lambda_0 = ck_0$ is the corresponding frequency which will turn out to be at the center of the photonic band gap (mid-gap). The detailed properties of this unit cell have been worked out using matrix transfer methods in previous works [9, 13], so I will just lift the needed formulas for this particular spontaneous emission problem.

First, let me define some relevant $n_i$-to-$n_j$ interface Fresnel transmission and reflection coefficients as,

$$t_{ij} = \frac{2n_i}{n_i + n_j} \quad , \tag{20a}$$

$$r_{ij} = -\frac{n_i - n_j}{n_i + n_j} = -r_{ji} \quad , \tag{20b}$$

where, using these, I can define two, special, double-boundary transmittance and reflectance coefficients as $T_{ij} = t_{ij}\,t_{ji}$, and $R_{ij} = -r_{ij}\,r_{ji}$, which have the energy conserving property that $T_{ij} + R_{ij} = 1$. Now, for the solution of the spontaneous emission problem in the $N$-period stack, I need the complex transmission and reflection coefficients, $t$ and $r$, for the quarter-wave unit cell in Fig. (3a). These are [9, 13],

$$t = \frac{T_{12}\,e^{i\pi\tilde{\omega}}}{1 - R_{12}e^{i\pi\tilde{\omega}}} \quad , \tag{21a}$$

$$r = \frac{r_{12}\left(e^{2i\pi\tilde{\omega}} - e^{i\pi\tilde{\omega}}\right)}{1 - R_{12}e^{i\pi\tilde{\omega}}} \quad , \tag{21b}$$

where $\tilde{\omega} = \omega/\omega_0$ is a dimensionless frequency normalized to that at midgap. Inserting these expressions, Eqs. (21), into Eq. (15) for the $N$-period DOM, gives for the quarter-wave (QW) stack [9, 13],

$$\rho_N^{\text{QW}}(\omega) = \frac{\pi T_{12}}{\omega_0 Nd}\,\frac{2\,N\,T_{12}\cos^2\left(\pi\tilde{\omega}/2\right) + R_{12}\,\sin^2\left(\pi\tilde{\omega}/2\right)\Xi_{2N}(\beta)}{\left(1 - 2R_{12} + \cos\pi\tilde{\omega}\right)\left[T_{12}^2\Theta_N^2(\beta) + \sin^2\pi\tilde{\omega}\,\Xi_N^2(\beta)\right]} \quad , \tag{22}$$

where $\beta = \arccos\left[\left(\cos\pi\tilde{\omega} - R_{12}\right)/T_{12}\right]$.

I plot this quarter-wave stack DOM, Eq. (22), in Fig. 4 for the parameters $N=5$, $n_1=1$, $n_2=2$, and $\omega_0=1$. I have chosen for convenience to normalize the DOM to $1/\rho_{bulk} = v_{bulk} = c(1/n_1 + 1/n_2)/2$, which is the harmonic mean of the group velocities in the two regions of the quarter-wave, bi-layer unit cell [9]. The frequency $\omega=\omega_0$ corresponds to the middle of the photonic band gap. I draw only first half of the gap, since for a quarter-wave stack the DOM is symmetric about midgap [9,13]. The wiggles in the density of modes in the pass band correspond to transmission resonance frequencies where the transmission of the stack $T_N = 1$. These resonance points of the $N$-period PBG stack correspond to the Fabry-Pérot resonances of a virtual dielectric slab of length $D = Nd$ whose effective index of refraction obeys the dispersion relation $\varphi_N = k_N(\omega)D = n_N(\omega)k_0 D$, where $\varphi_N$ is the phase of the $N$-period transmission coefficient $t_N$, the factor $n_N(\omega)$ is the effective index of the virtual slab, and $k_0 = \omega_0/c$ is the vacuum wave number with $c$ the vacuum speed of light as usual.

Now the electric field inside the $n^{th}$ unit cell, Fig. (3b), must be written as function that depends on whether the field point $x$ is in the $n_1$ or the $n_2$ region. Hence, one can write,

$$E_n^I(\omega,x) = A_n(\omega)\cos\left(\frac{\pi}{2}\tilde{\omega}\,\frac{x}{a}\right) + B_n(\omega)\sin\left(\frac{\pi}{2}\tilde{\omega}\,\frac{x}{a}\right) , \quad \{(n-1)d \leq x < (n-1)d + a\} \qquad (23a)$$

$$E_n^{II}(\omega,x) = C_n(\omega)\cos\left(\frac{\pi}{2}\tilde{\omega}\,\frac{x}{b}\right) + D_n(\omega)\sin\left(\frac{\pi}{2}\tilde{\omega}\,\frac{x}{b}\right) , \quad \{nd - b \leq x < nd\} \qquad (23b)$$

for the $n^{th}$ unit cell field. Now I may apply the result of Eq. (17), or directly use the BC in Eq. (16)—together with the demand that the field be continuous and differentiable at the $n_1$ to $n_2$ interface, to obtain,

$$A_n = \bar{\lambda}_n \ , \qquad (24a)$$

$$B_n = \left[(\delta n - \cos\pi\tilde{\omega})\bar{\lambda}_n + (1-\delta n)\bar{\rho}_n\right]\csc\pi\tilde{\omega} \ , \qquad (24b)$$

$$C_n = \left\{\left[(1+\delta n)\bar{\lambda}_n - \delta n\bar{\rho}_n\right]\sin\left(\frac{\pi\tilde{\omega}}{1+\delta n}\right) + \bar{\rho}_n\sin\left(\frac{\delta n\pi\tilde{\omega}}{1+\delta n}\right)\right\}\csc\pi\tilde{\omega} \ , \qquad (24c)$$

$$D_n = \left\{\left[\delta n\bar{\rho}_n - (1+\delta n)\bar{\lambda}_n\right]\cos\left(\frac{\pi\tilde{\omega}}{1+\delta n}\right) + \bar{\rho}_n\cos\left(\frac{\delta n\pi\tilde{\omega}}{1+\delta n}\right)\right\}\csc\pi\tilde{\omega} \ , \qquad (24d)$$

where I have defined $\bar{\lambda}_n \equiv \lambda_n^{(1)} + \lambda_n^{(2)}$ and $\bar{\rho}_n \equiv \rho_n^{(1)} + \rho_n^{(2)}$ in terms of sums of the components of the $n^{th}$ unit cell boundary-value vectors, Eqs. (16). The differential refractive index factor by $\delta n = (n_1 - n_2)/(n_1 + n_2) = -r_{12}$. Using the complex quarter-wave transmission and reflection coefficients, Eqs. (20), I obtain,

$$\bar{\rho}_n = \frac{T_{12}^2 e^{2i\pi\tilde{\omega}}\Xi_{N-n-1} - \left[r_{12}^2 e^{2i\pi\tilde{\omega}}(1-e^{\pi i\tilde{\omega}})^2 + (1+e^{\pi i\tilde{\omega}}R_{12})^2\right]\Xi_{N-n}}{T_{12}^2 e^{2i\pi\tilde{\omega}}\Xi_{N-1} - (1+R_{12}e^{\pi i\tilde{\omega}})^2\Xi_N} \ , \qquad (25a)$$

$$\bar{\lambda}_n = \frac{T_{12}^2 e^{2i\pi\tilde{\omega}}\Xi_{N-n} - \left[r_{12}^2 e^{2i\pi\tilde{\omega}}(1-e^{\pi i\tilde{\omega}})^2 + (1-e^{\pi i\tilde{\omega}}R_{12})^2\right]\Xi_{N-n+1}}{T_{12}^2 e^{2i\pi\tilde{\omega}}\Xi_{N-1} - (1-R_{12}e^{\pi i\tilde{\omega}})^2\Xi_N} \ , \qquad (25b)$$

for the boundary condition quantities. Inserting Eqs. (25) into Eqs. (24) into Eqs. (23) gives us the analytic solution for $E_n(\omega,x)$ in the $n^{th}$ unit cell.

As an example, in Fig. 5 I plot the intensity $I = |E_3(\omega,x)|^2$ in the middle-most unit cell ($n = 3$), for an $N = 5$ period stack. I again use the index values of $n_1 = 1$ and $n_2 = 2$ for illustration. In Fig. 5a the intensity in the $n_1$ layer is plotted over the range $x \in [(n–1)d, (n–1)d+a]$. In Fig. 5b I have the intensity in the $n_2$ layer over the range $x \in [nd–b, nd]$. (In both cases I have translated the unit cell to a new origin and scaled by the dielectric thickness.) The plots are represented as a 2D function of both dipole frequency $\omega$ and position $x$. The electric modal nodal and antinodal behavior at different frequencies is clearly visible. In particular, notice that at the low-frequency band edge, at around $\omega/\omega_0$=0.75, the low-index field $I_a$ experiences an antinode, while the intensity $I_b$ in the high-index region exhibits a node. This is a well-known property of the field modes at the low-frequency photonic band edge [9, 11, 22]. This behavior is reversed at the high-frequency edge, which is not displayed here. The other pronounced modal distributions are at the transmission resonance frequencies, which correspond to Fabry-Pérot oscillations in a virtual dielectric slab of effective index $n_N(\omega)$ as discussed above.

These modal field intensities, when normalized to the modal energy as per Eqs. (18) and (19), provide the final missing piece for the solution to the spontaneous emission formula, Eq. (4). This equation can be written in 1D notation as,

$$p_n(\omega, x) = \rho(\omega) \, |e_n(\omega, x)|^2 , \qquad (26)$$

which yields the scaled emitted power for a normal dipole (perpendicular to the $x$ direction), of frequency $\omega$ and position $x$ inside the $n^{th}$ unit cell (I have dropped the subscripts naught on $\omega$ and $x$). The full expression is too large to be presented here; it is more useful now to plot the results for the specific example I have been using with the central unit cell $(n = 3)$, of an $N = 5$ period stack. I again use the index values of $n_1 = 1$ and $n_2 = 2$.

First, using the modal fields above, I must calculate the normalization function $U_5(\omega)$ given by Eq. (18). The result for the example five-period stack is plotted in Fig. (6). Note that there are four resonances in the pass band, corresponding to the four transmission resonances seen in the DOM, Fig. (4). Again, these resonances can be thought of as Fabry-Pérot resonances of a virtual dielectric slab with an effective index of refraction equal to that of the PBG stack, as discussed above. It is well known that the DOM is very large at such resonances where the structure behaves like a high-$Q$ cavity, and hence this is where the energy stored in the structure is large when compared to other frequencies, as is clear in the plot. Once this energy normalization function is established, I can then generated the final result of the 1D emission rate $p_3(\omega, x)$, given in Eq. (26), for a point dipole emitter of frequency $\omega$ and position $x$ inside the third (middle) unit cell of the five-period quarter-wave stack that I have been using throughout. The results for the $n_1 = 1$ and $n_2 = 2$ regions are plotted separately in Figs. 7a and 7b, respectively. I have chosen the dipole to be located in the geometric center of each layer, which in my translated and scaled units corresponds to $x = a/2$ and $x = b/2$, respectively. Such a choice corresponds qualitatively to the average response of a collection of dipoles distributed uniformly throughout the layer [11]. We can see the effect of the band-gap $(\omega > 0.75\omega_0)$ is to suppress the emission for both types of layer as expected. At the low-frequency band-edge resonance near $\omega = 0.75\omega_0$, we see in Fig. 7a that the low-index emission is relatively suppressed even though the density of modes, Fig. (4), is very large here. This result is due to the fact that the low-index modal field exhibits a node at the dipole location, Fig. 5a, and so the radiation-reaction and vacuum fields couple very weakly to the dipole in spite of the large DOM. The opposite is true in the high-index layer where the dipole sits at a modal node, Fig. 5b, and so the fields couple very strongly to the dipole, enhancing the emission, as seen in Fig. 7b. This band-edge emission enhancement was seen both numerically experimentally by my former co-workers and me in a 1D active semiconductor stack [11]. My exact model presented here agrees qualitatively quite well with those results.

## 4. SUMMARY AND CONCLUSIONS

In this work I have presented, for the first time, a complete solution of the point dipole emission problem in a simple model of a *finite,* 1D, photonic band-gap material. The study of spontaneous emission in cavities has a long tradition of theory [23] and experiment [24] in the field of cavity QED. A photonic band-gap structure is particular type of dielectric cavity where interference effects give rise to the cavity confinement and the alteration of the electromagnetic modal structure. The fact that a finite, 1D PBG material is quasi-periodic allows one to make quantitative statements about the density of modes and the modal eigenfunctions, using some simple analytical results from matrix transfer theory. In particular, in Sec. 2 I give a formula for the scaled dipole emission rate, Eq. (4), in terms of the density of modes and the electromagnetic eigenmode function. For a general 1D PBG stack composed of arbitrary, lossless, dispersionless unit cells, I give the Scattering Matrix Reduction Formula, Eq. (13), which allows one to express the transfer matrix for the entire stack (or any subset thereof) in terms of the matrix for the unit cell and the Bloch phase. With this in hand, an exact solution of the density of modes can be found, Eq. (15), as well as the modal fields, Eqs. (17), giving the 1D emission rate. To illustrate these ideas, in Sec. 3 I give a specific example of the emission in the case of a quarter wave stack, for which the density of modes is given by Eq. (22) and the electric field modes by Eqs. (23) and (24). Using these results, I am able to calculate and plot the density of modes, electric field intensity, total electromagnetic energy, and finally the point dipole emission rate, all in a five-period, quarter-wave stack. The result of this work has many applications to the study of wave phenomena in 1D periodic structures. Although this paper was designed with electromagnetic waves in mind, the result is very general and can apply to electron matter waves in semiconductor superlattices [25], sound wave or phonon emission in periodic structures [26], and atomic matter-waves in 1D optical molasses periodic potentials [27].

## ACKNOWLEDGEMENTS

The research described in this paper was carried out by the Jet Propulsion Laboratory, California Institute of Technology, under a contract with the National Aeronautics and Space Administration.

## FIGURE CAPTIONS

Fig. 1    Scattering an electromagnetic wave off of a finite, 1D, inhomogeneous dielectric. An electromagnetic wave is incident from the left and normalized to have amplitude unity and zero phase at the leftedge of the potential. The index $n(x)$ varies with position $x$ over the range $(0, d)$ but is dispersionless. The complex transmission and reflection coefficients $t$ and $r$ can be calculated once the form of $n(x)$ is specified. A point dipole of moment $\mu$, oriented perpendicular to the $x$ axis, is located at a point inside the dielectric.

Fig. 2    The index profile in Fig. 1 is now repeated periodically $N$ times to form a finite, 1D, $N$-period, photonic band structure. The stack transmission and reflection coeficients $t_N$ and $r_N$ can be found in terms of the unit cell quantities $t$, $r$, and the Bloch phase $\beta$.

Fig. 3    In (a) I depict a unit cell scatterer for making a quarter-wave stack. The unit cell has two layers of indices $n_1$ and $n_2$. The thickness of each layer is chosen to be a quarter of some reference wavelength $\lambda_0$. The cell is sandwiched between two semi-infinite regions of index $n_1$. In (b) this quarter-wave cell is repeated $N$ times to form an $N$-period quarter-wave stack. The emission rate of a point dipole emitter in the $n^{th}$ unit cell is derived.

Fig. 4    The density of modes (DOM), Eq. (22), is plotted here as a function of the dimensionless frequency $\omega/\omega_0$ for a five-period quarter wave stack with layer indices $n_1 = 1$ and $n_2 = 2$. I have normalized the DOM to a bulk velocity $v_{bulk}$, which corresponds to scaling the DOM by that of an infinite homogeneous material with an effective index that is the harmonic mean of $n_1 = 1$ and $n_2 = 2$. The photonic band gap is in a region $\omega/\omega_0 \in [0.75, 1]$. (The DOM of a quarter-wave stack is symmetric about the point $\omega/\omega_0 = 1$.) The supression of the DOM in the gap is clear, as is the enhancement of the DOM at the band-edge resonance at $\omega/\omega_0 = 0.75$, where the transmission of the stack is unity. There are other less-pronounced peaks in the DOM in the pass band at frequencies that correspond to other transmission resonances.

Fig. 5    Depicted here are the unnormalized electric field intensities calculated from Eqs. (23) (in arbitrary dimensionless units) in the third (middle) unit cell of a five-period quarter-wave stack. In (a) I show the intensity in the $n_1$ layer and in (b) the $n_2$ layer. These intensities are given as a function of both scaled frequency $\omega/\omega_0$ and the scaled position $x/a$ or $x/b$ inside the $n_1 = 1$ and $n_2 = 2$ layers in (a) and (b), respectively. The position of the origin at $x = 0$ is with respect to a coordinate system translated up to the unit cell layer. The suppression of the field in the band gap region, $\omega/\omega_0 \in [0.75, 1]$, is clear in both (a) and (b). The field in the $n_1$ (low-index) region at the photonic band edge, $\omega/\omega_0 = 0.75$, exhibits a spacial node, while that in the $n_2$ (high-index) region shows an antinode. This feature is a well-known property of the modal fields at the low-frequency edge of the photonic band gap. (Field intensities are not symmetric in each layer due to finite-size effects.) The electric field oscillations at the pass-band frequencies correspond to other transmission resonance Fabry-Pérot oscilla-

tions of a virtual dielectric slab having an effective index given by the dispersion relation of the stack.

Fig. 6    The total electric field mode energy, Eq. (18), of the entire five-period quarter wave stack is plotted here in arbitrary dimensionless units versus scaled frequency $\omega/\omega_0$. The stack acts like a Fabry-Pérot oscillator at the pass-band transmission frequencies, as described in the text, giving rise to the series of energy peaks in the pass band. The energy stored in the stack is suppressed in the stop band, $\omega/\omega_0 \in [0.75, 1]$, as expected.

Fig. 7    The scaled emission rate, Eq. (26), of a dipole located in the center of the $n_1 = 1$ and $n_2 = 2$ regions is plotted in (a) and (b), respectively. The emission is proportional to the product of the density of modes, Fig. 4, and the electric field mode intensities, Fig. 5. Hence, in the band gap, $\omega/\omega_0 \in [0.75, 1]$, the emission is strongly suppressed, as expected, since the modal functions and the DOM are both suppressed here. In the $n_1 = 1$ region (a), the emission is suppressed at the band-edge frequency, $\omega/\omega_0 = 0.75$, because the dipole is located at a node, even though the DOM is large. Contrariwise, in the $n_2 = 2$ region (b) the band-edge emission is enhanced because both the DOM is large and the dipole is at an anti-node and so couples well with the field mode. Other peaks in the pass band correspond to Fabry-Pérot modes of an effective dielectric stack and they reflect both the DOM and the local modal structure.

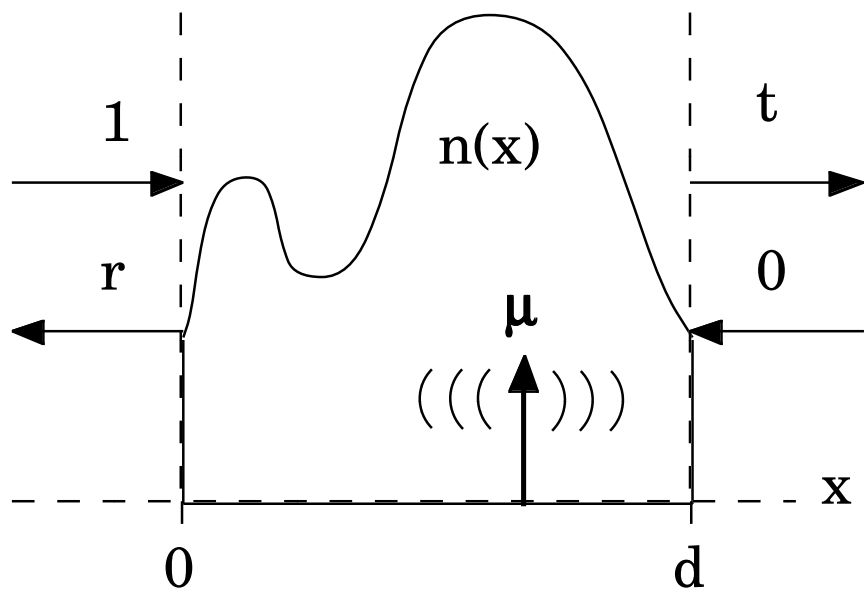

Fig. 1

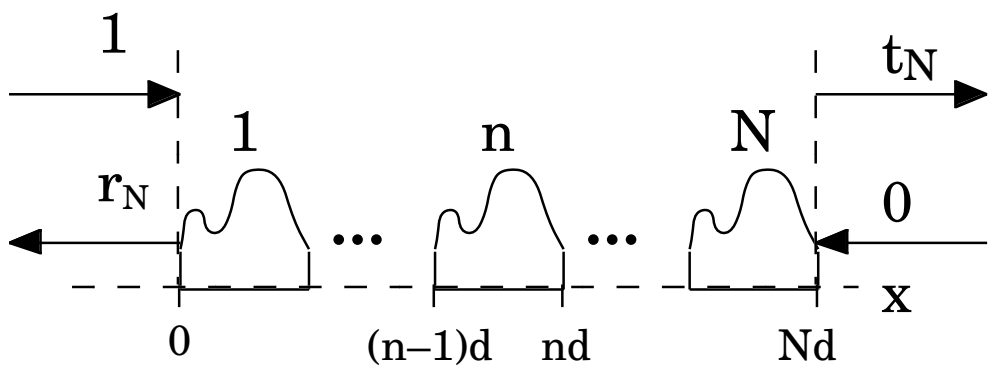

Fig. 2

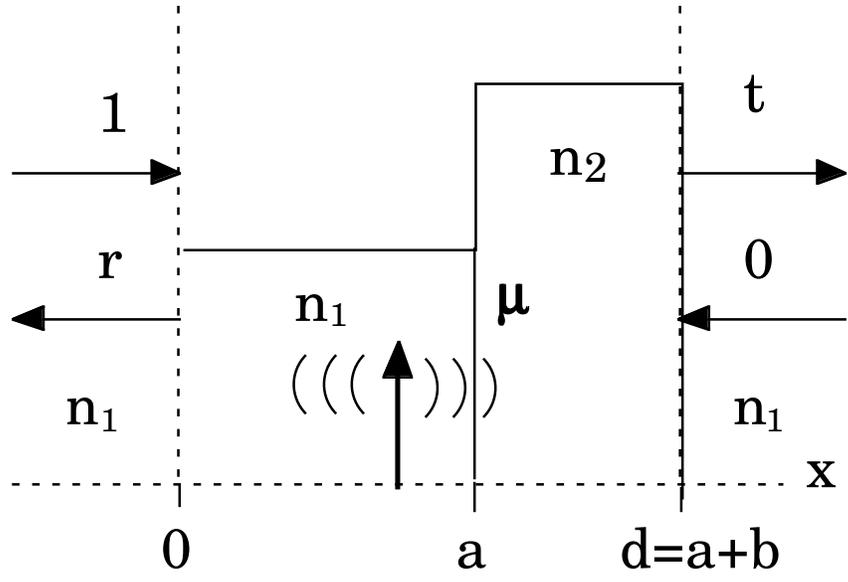

(a)

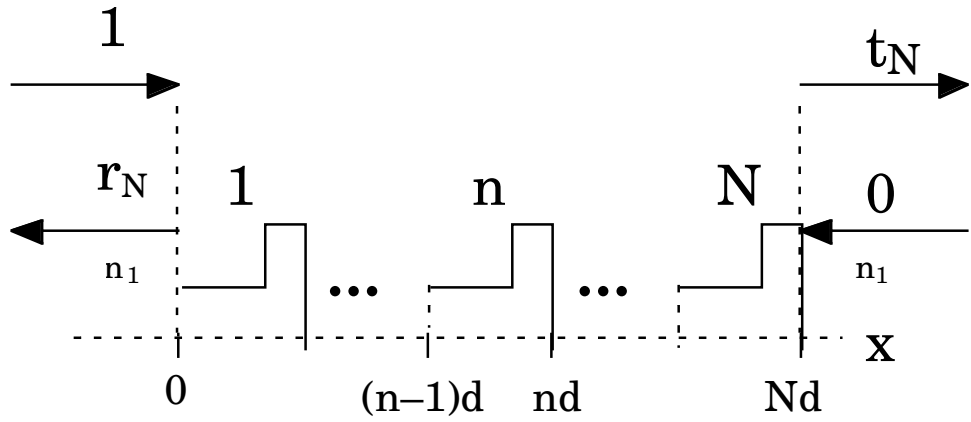

(b)

Fig. (3)

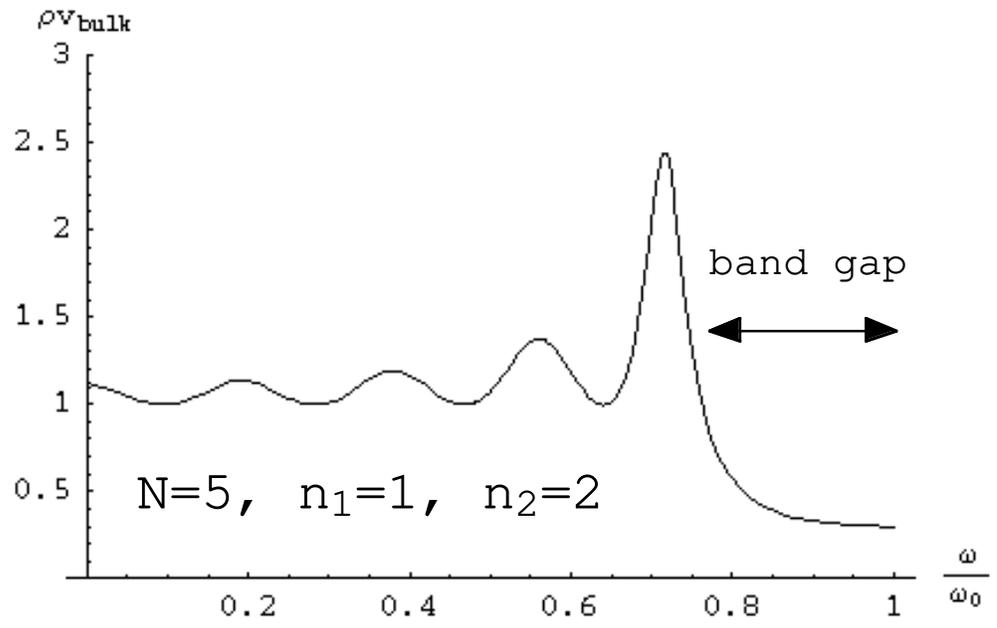

Fig. (4)

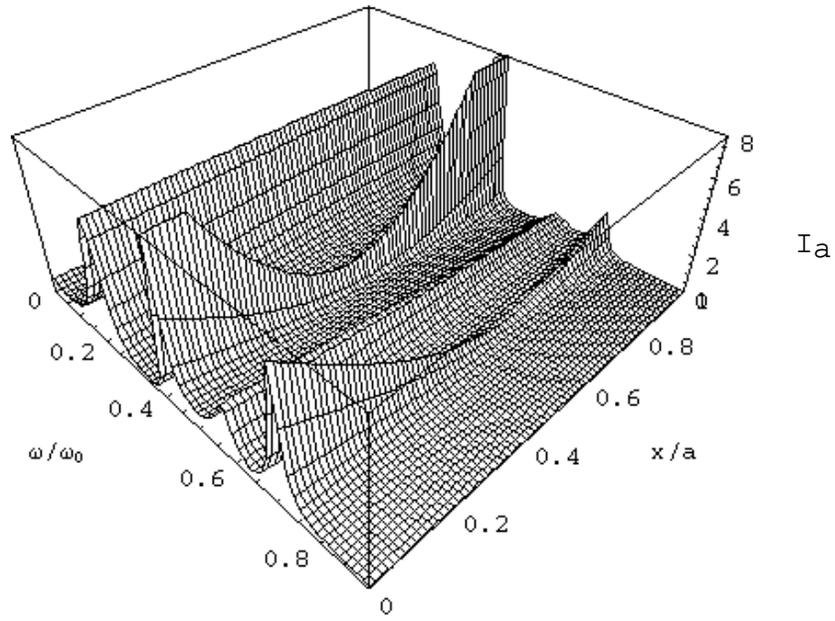

(a)

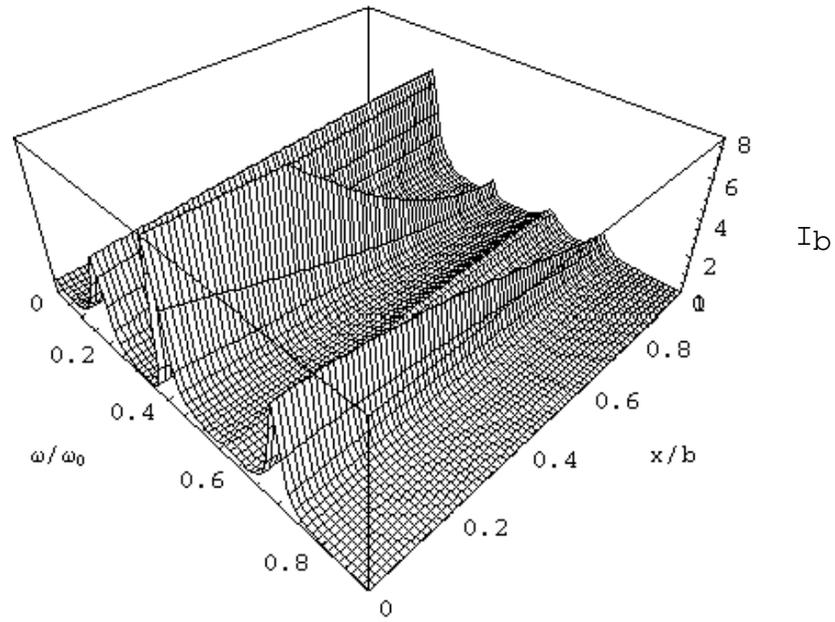

(b)

Fig. (5)

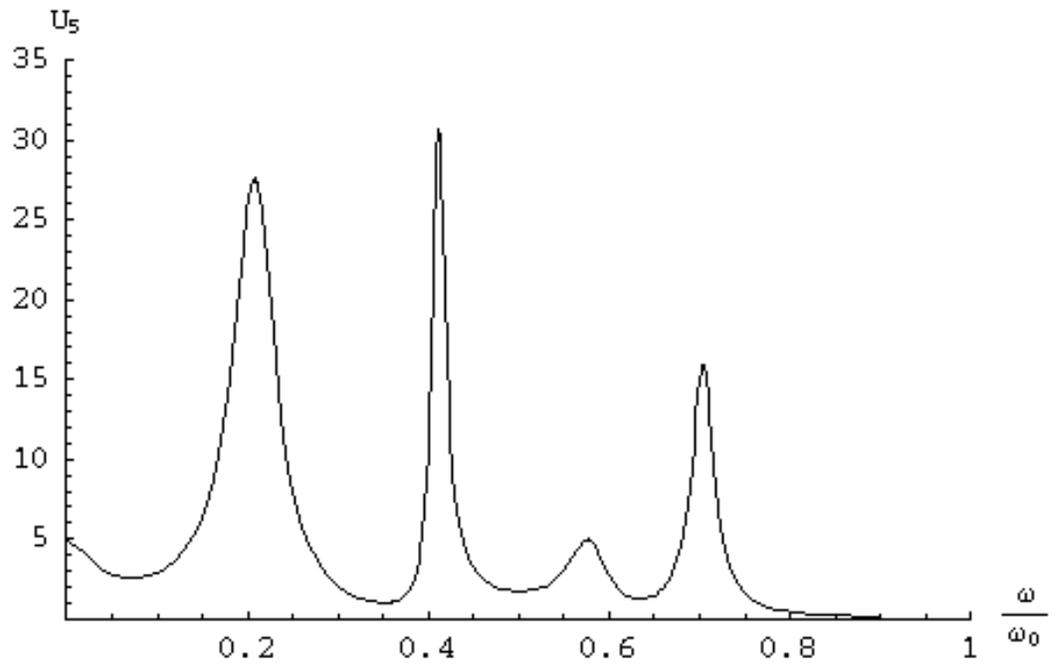

Fig. (6)

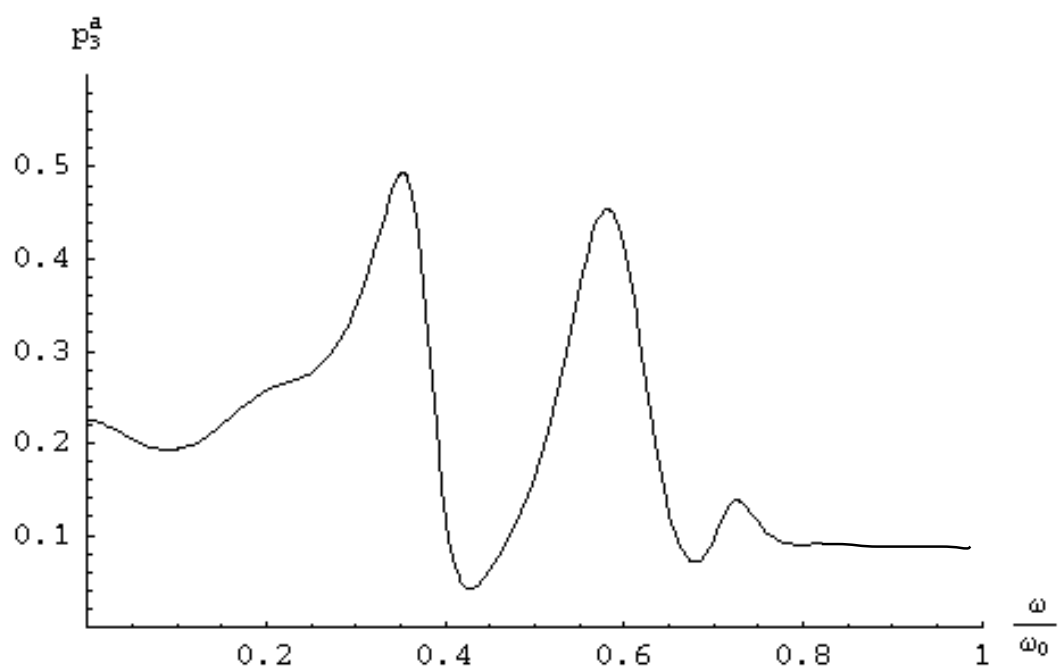

(a)

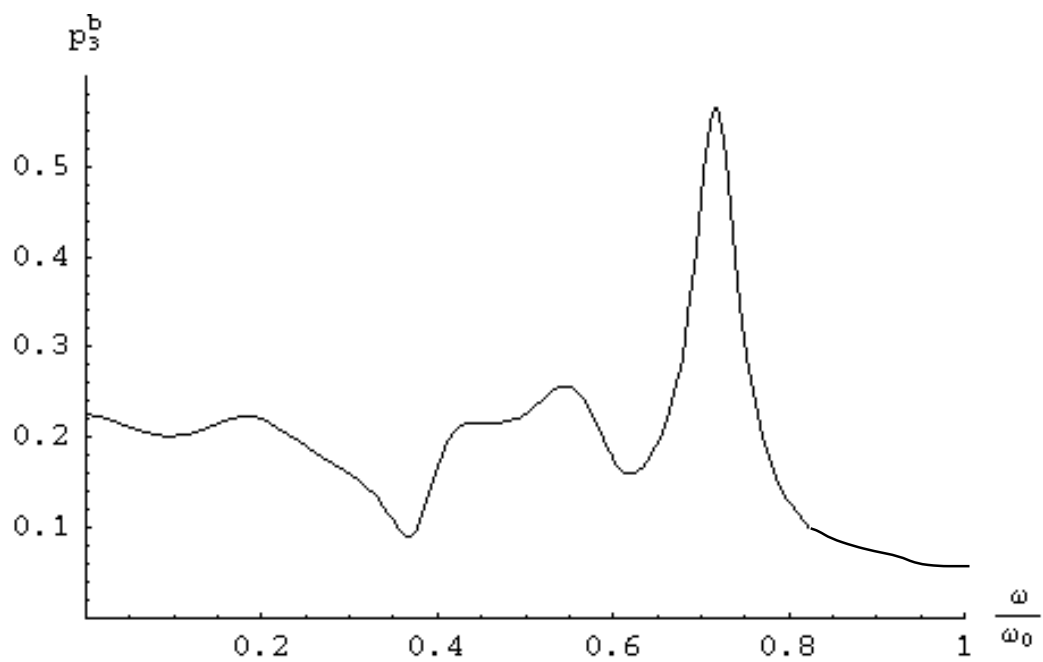

(b)

Fig. (7)